\documentclass{webofc}
\usepackage[varg]{txfonts}   

\usepackage{slashed}
\usepackage{graphicx}
\usepackage{subfigure}
\usepackage{soul}
\usepackage{amsmath}
\usepackage{multirow}
\usepackage{tablefootnote}
\usepackage{hyperref}
\usepackage{epstopdf}
\usepackage{lscape}
\usepackage{url}
\usepackage[dvipsnames]{xcolor}
\usepackage{xcolor}
\usepackage{hyperref}
\usepackage{fancyhdr}

\def\figureautorefname~#1\null{Fig.\,#1\null}
\def\tableautorefname~#1\null{Tab.\,#1\null}
\def\equationautorefname~#1\null{Eq.\,(#1)\null}

\begin{document}
%

\title{Exploring the 2HDM with Global Fits in \texttt{GAMBIT}}
%
%

\author{
\firstname{Filip } 
\lastname{Rajec}
\inst{1}\fnsep\thanks{\email{filip.rajec@adelaide.edu.au}} \and
\firstname{Wei} \lastname{Su}\inst{1}
\fnsep\thanks{Speaker, \email{wei.su@adelaide.edu.au}} \and
\firstname{Martin} \lastname{White}\inst{1}
\fnsep\thanks{\email{martin.white@adelaide.edu.au}} \and
\firstname{Anthony G. } \lastname{Williams}\inst{1}
\fnsep\thanks{\email{anthony.williams@adelaide.edu.au}}
}
\institute{ARC Centre of Excellence for Particle Physics at the Terascale, \\Department of Physics,University of Adelaide, South Australia 5005, Australia }

\abstract{%
  In this work, we present \emph{preliminary} results of a global fit of the type-II two-Higgs-doublet model (2HDM) with the tool \texttt{GAMBIT}.  Our study includes various constraints, including the theoretical constraints (unitarity, perturbativity and vacuum stability), Higgs searches at colliders, electroweak physics and flavour constraints.  With the latest experimental results, our results not only confirm past studies but also go further in probing the model.  We find, for example, that the measurements of $B\rightarrow K^*\mu^+\mu^-$ angular observables cannot be explained in the type-II 2HDM.
}
\maketitle
\thispagestyle{fancy}

\section{Introduction}
\label{intro}

The discovery of a Standard Model (SM)-like Higgs boson at the Large Hadron Collider (LHC)~\cite{Aad:2012tfa,Chatrchyan:2012xdj} strongly motivates LHC searches for physics beyond-the-SM (BSM).  Models with extended Higgs sectors provide promising candidates for this new physics.

In this study, we carry out global fits of the $Z_2$-Yukawa symmetric Two-Higgs-Doublet Model (2HDM) \cite{Branco:2011iw}, specifically the type-I, type-II, lepton specific and flipped models.  This analysis is carried out by the open-source tool \texttt{GAMBIT} \cite{Athron_2017} (Global and Modular beyond-Standard Model Inference Tool). In these proceedings, we present \emph{preliminary} results, discussing the constraints on the type-II 2HDM. We include theoretical constraints (unitarity, perturbativity and vacuum stability), Higgs searches at colliders, electroweak physics and flavour constraints.

In Section~\ref{sec:2hdm}, we briefly introduce the 2HDM, and present our results in Section~\ref{sec:res}. We offer our conclusions in Section~\ref{sec:con}.

\section{The Two-Higgs-Doublet Model}
\label{sec:2hdm}

2HDMs are encountered in various attempts to solve the problems of the Standard Model, including the Minimal Supersymmetric Standard Model, gauge extensions (such as the Left-Right symmetric model), and flavour models. Thus, exploring the physics of 2HDMs with the latest experimental constraints provides unique information covering a broad class of BSM scenarios.

A general 2HDM introduces two ${\rm SU}(2)_L$ scalar doublets $\Phi_i$, $i=1,2$,

\begin{equation}
\Phi_{i}=\begin{pmatrix}
  \phi_i^{+}    \\
  (v_i+\phi^{0}_i+iG_i)/\sqrt{2}
\end{pmatrix}.
\end{equation}
Each obtains a VEV $v_1$ or $v_2$ after electroweak symmetry breaking (EWSB) with $v_1^2+v_2^2 = v^2 = (246\ {\rm GeV})^2$, and $v_2/v_1=t_\beta$\footnote{Throughout this work we use the notation $c_x$, $s_x$ and $t_x$ to refer to $\cos(x), \sin(x)$ and $\tan(x)$ respectively. Specifically for the angle combination $\beta-\alpha$ we write $c_{\beta\alpha}$ and $s_{\beta\alpha}$.}.

The 2HDM Lagrangian for 
the Higgs sector can be written as
\begin{equation}\label{}
\mathcal{L}=\sum_i |D_{\mu} \Phi_i|^2 - V(\Phi_1, \Phi_2) + \mathcal{L}_{Yuk},
\end{equation}
 with the Higgs potential  
 \begin{eqnarray}
 V(\Phi_1, \Phi_2)=&& m_{11}^2\Phi_1^\dag \Phi_1 + m_{22}^2\Phi_2^\dag \Phi_2 -m_{12}^2(\Phi_1^\dag \Phi_2+ h.c.) + \frac{\lambda_1}{2}(\Phi_1^\dag \Phi_1)^2 + \frac{\lambda_2}{2}(\Phi_2^\dag \Phi_2)^2  \notag \\
 & &+ \lambda_3(\Phi_1^\dag \Phi_1)(\Phi_2^\dag \Phi_2)+\lambda_4(\Phi_1^\dag \Phi_2)(\Phi_2^\dag \Phi_1)+\frac{1}{2}\Big[ \lambda_5(\Phi_1^\dag \Phi_2)^2 + h.c.\Big],
\end{eqnarray}
assuming CP-conservation and a soft ${Z}_2$ symmetry breaking term $m_{12}^2$.

After EWSB, one of the four neutral components and two of the four charged components are eaten by the SM $Z$, and $W^\pm$, providing the gauge boson masses.  The remaining physical mass eigenstates are the two CP-even Higgses $h$ and $H$, one CP-odd Higgs $A$ and a pair of charged Higgs bosons $H^\pm$. Here we take $m_h<m_H$. Usually we take the general basis of the eight parameters appearing in the Higgs potential: $(m_{11}^2, m_{22}^2, m_{12}^2, \lambda_{1,2,3,4,5})$. A more convenient choice (the physical basis) is given by: $(v, t_\beta, \alpha, m_h, m_H, m_A, m_{H^\pm}, m_{12}^2)$, in which  $\alpha$ is the rotation angle diagonalizing the CP-even Higgs mass matrix.
\rhead{}
\section{Study Results}
\label{sec:res}
\texttt{GAMBIT} is compatible with both the Bayesian and frequentist statistical frameworks, and we here focus on frequentist results obtained with the \texttt{Diver} implementation of the differential evolution algorithm ~\cite{Workgroup:2017htr}. Our results converged well with NP= 5000, convthresh=1e-5. The 2HDM parameters are explored with the ranges $\lambda_{1,2,3,4,5}\in(-3\pi,3\pi)$, $\lambda_{6,7}=0$, $m_{12}^2 \in (- 10^6, 10^7) ~\rm{GeV}^2$
, and a combined likelihood for each parameter point is calculated by combining individual likelihoods for a variety of experimental and theoretical constraints. Theoretical constraints ensure that our potential is bounded from below, our vacuum is stable at tree-level, and that the scattering amplitude eigenvalues give a unitary $S$-matrix (up to NLO).  We check that our potential is bounded from below and our four-Higgs couplings (equivalent to tree-level scattering eigenvalues) are perturbative up to 1 TeV.

Experimentally, we combine observations from LEP, ATLAS and CMS, flavour physics observations and also measurements of the electroweak precision parameters. 
The 2HDM spectrum is generated at two-loops using FlexibleSUSY. 
Final results are presented in form of 1D and 2D profile log-likelihood distributions for various parameter combinations.  The $1\sigma$, 2$\sigma$ and 3$\sigma$ regions in these plots are calculated using the difference in profile log-likelihood ratio from the best fit point, assuming the validity of Wilks' theorem. This in turn relies on the assumption of a Gaussian likelihood. Our likelihood is Gaussian in the data itself, but our final results stem from performing a non-linear transformation from the data to the theory parameter space, which has been shown to lead to violations of the expected coverage of the $1\sigma$, 2$\sigma$ and 3$\sigma$ regions~\cite{Akrami:2010cz,Strege:2012kv}. Nevertheless, Wilks' theorem approximately holds, and we therefore follow the standard practise of presenting these regions as indicative approximations.

\begin{figure}[h]
\centering
\includegraphics[width=6cm,clip]{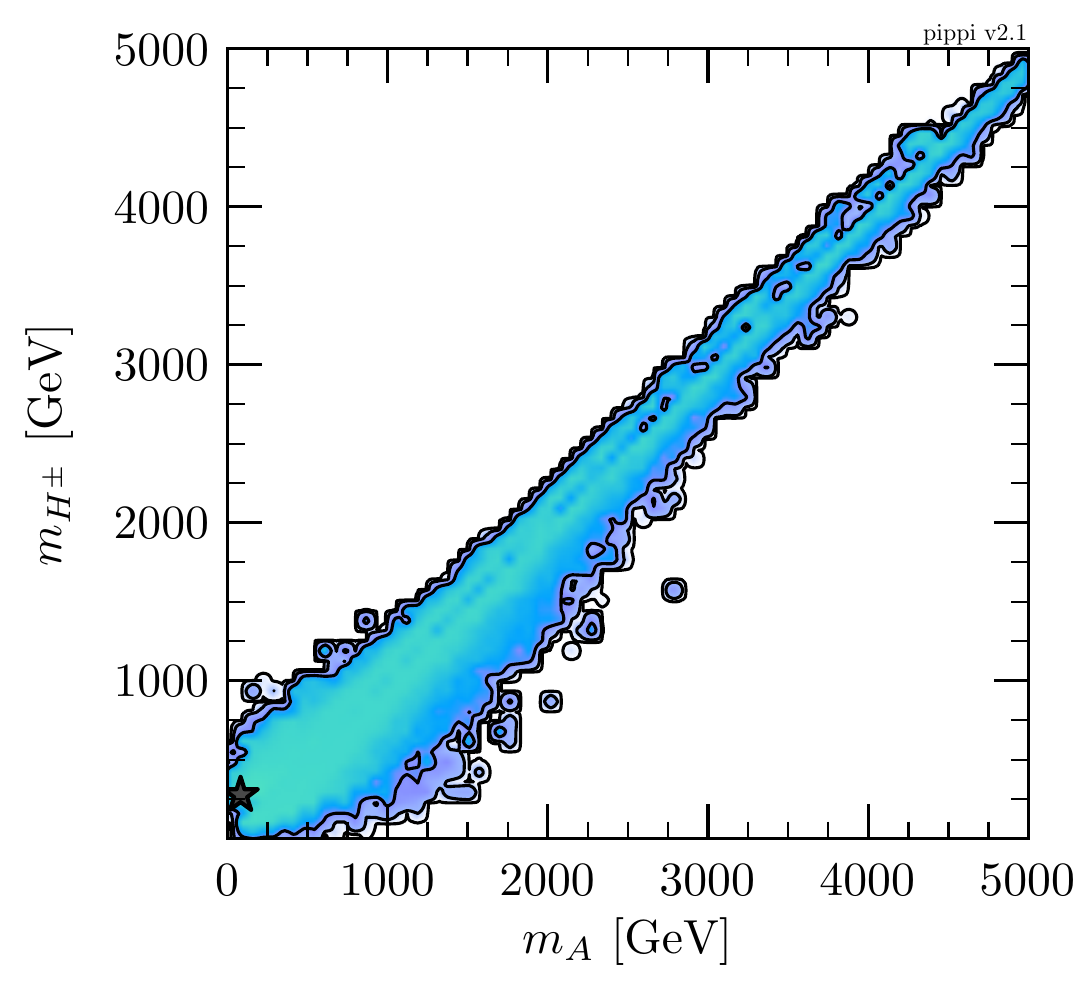}
\includegraphics[width=6cm,clip]{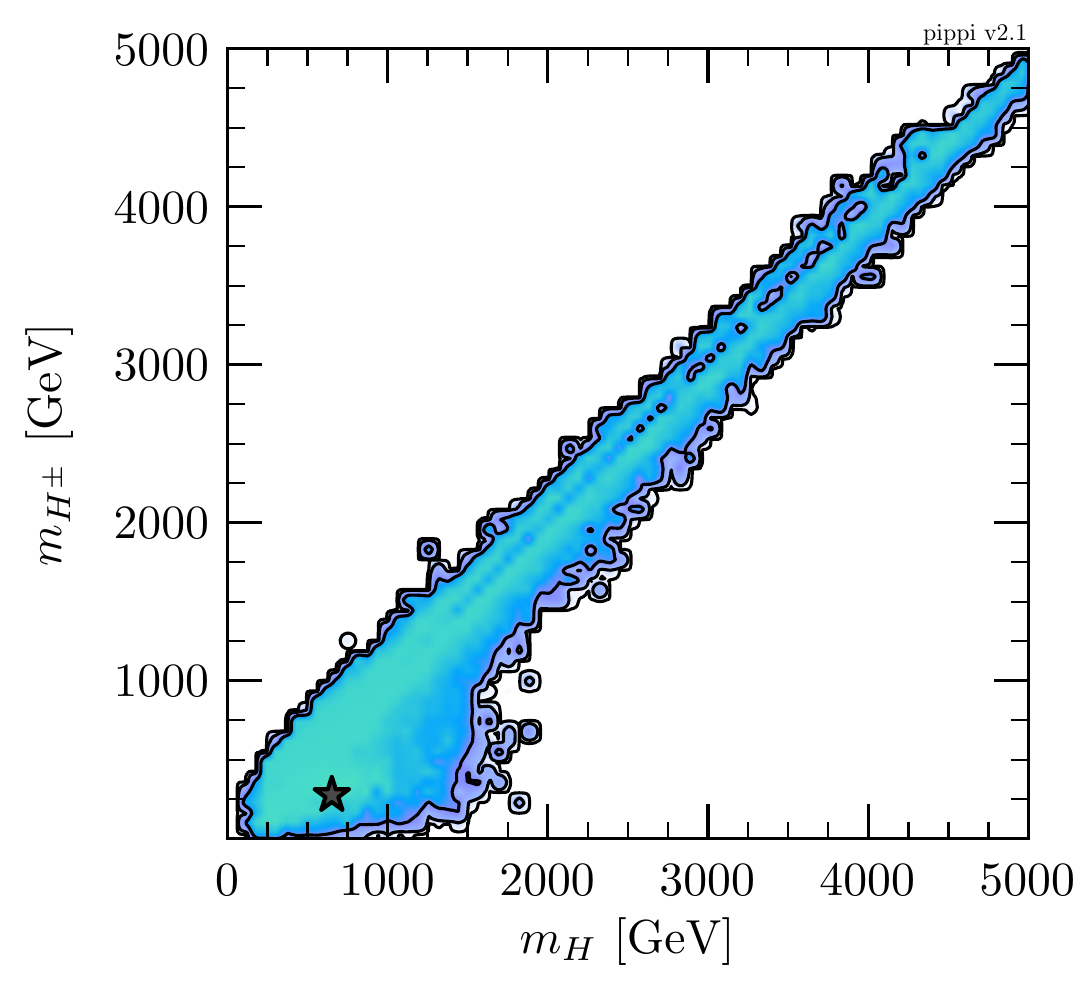}
\caption{2D profile likelihood distributions of electroweak precision measurements in the plane $m_H$ vs $m_{H^\pm}$ \emph{(left)} and $m_A$ vs $m_{H^\pm}$ \emph{(right)}. Generally the allowed region follows $m_{H^\pm} \approx m_A,m_H$. The solid lines represent $1,2$ and $3\, \sigma$ regions and the best-fit points are plotted as black stars. }
\label{fig:theory}       
\end{figure}
\subsection{Electroweak constraints}

Previous studies \cite{Kling:2016opi,Haber:2015pua,Chen:2019rdk,Chen:2018shg} have shown that the charged Higgs mass is constrained to be close to the mass of either of the neutral Higgses ($H$ or $A$) in order to satisfy the EW precision measurements. In \autoref{fig:theory}, we show the results of an electroweak parameters-only global fit in the plane $m_H$ vs $m_{H^\pm}$ on the left and $m_A$ vs $m_{H^\pm}$ on the right. Generally the allowed region locates $m_{H^\pm} \approx m_A/m_H$ up to 5 TeV.

\subsection{Theoretical constraints}

In this section, we will show the effects on the allowed 2HDM parameter space after imposing theoretical constraints. We impose unitarity, perturbativity and vacuum stability~\cite{Gu:2017ckc}.
To get a general idea, we adopt the simplification of $m_{H^{\pm}}=m_{H}=m_{A}\equiv m_{\phi}$ so that electroweak constraints are automatically satisfied. 

\begin{figure}[h]
\begin{center}
 \includegraphics[width=9cm]{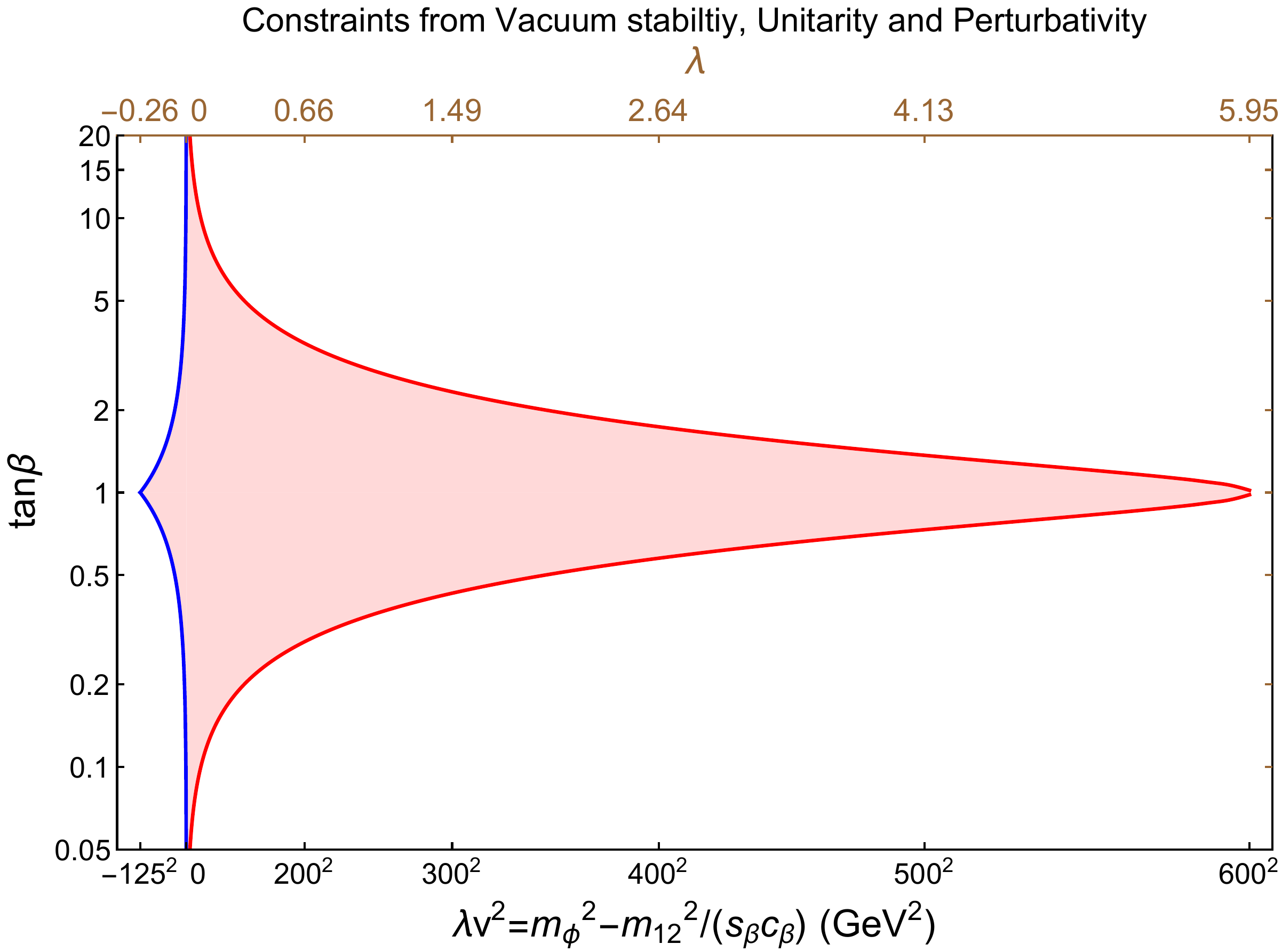}
\caption{The shaded region indicates the surviving region of 2HDM parameter space of  $t_\beta$ vs. $\lambda v^2$, after vacuum stability \emph{(blue lines)}, unitarity and perturbativity \emph{(red lines)} are taken into account at leading order.  The corresponding values of $\lambda$ are shown in the upper axis.  We assume that $m_{H^{\pm}}=m_{H}=m_{A}\equiv m_{\phi}$, and also that the alignment limit of $c_{\beta\alpha} = 0 $ is satisfied.}
\label{fig:constraint}
\end{center}
\end{figure}

We see that 

\begin{equation}
  -m_h^2 < \lambda v^2 < (600\  \text{GeV})^2,
     \label{eq:lam_cons}
\end{equation}
which gives $ -0.258 < \lambda =- \lambda_4=- \lambda_5  < 5.949$ and $0 <  \lambda_3  < 6.207$.
\subsection{Flavour constraints}

\begin{figure}[h]
\centering
\includegraphics[width=7cm,clip]{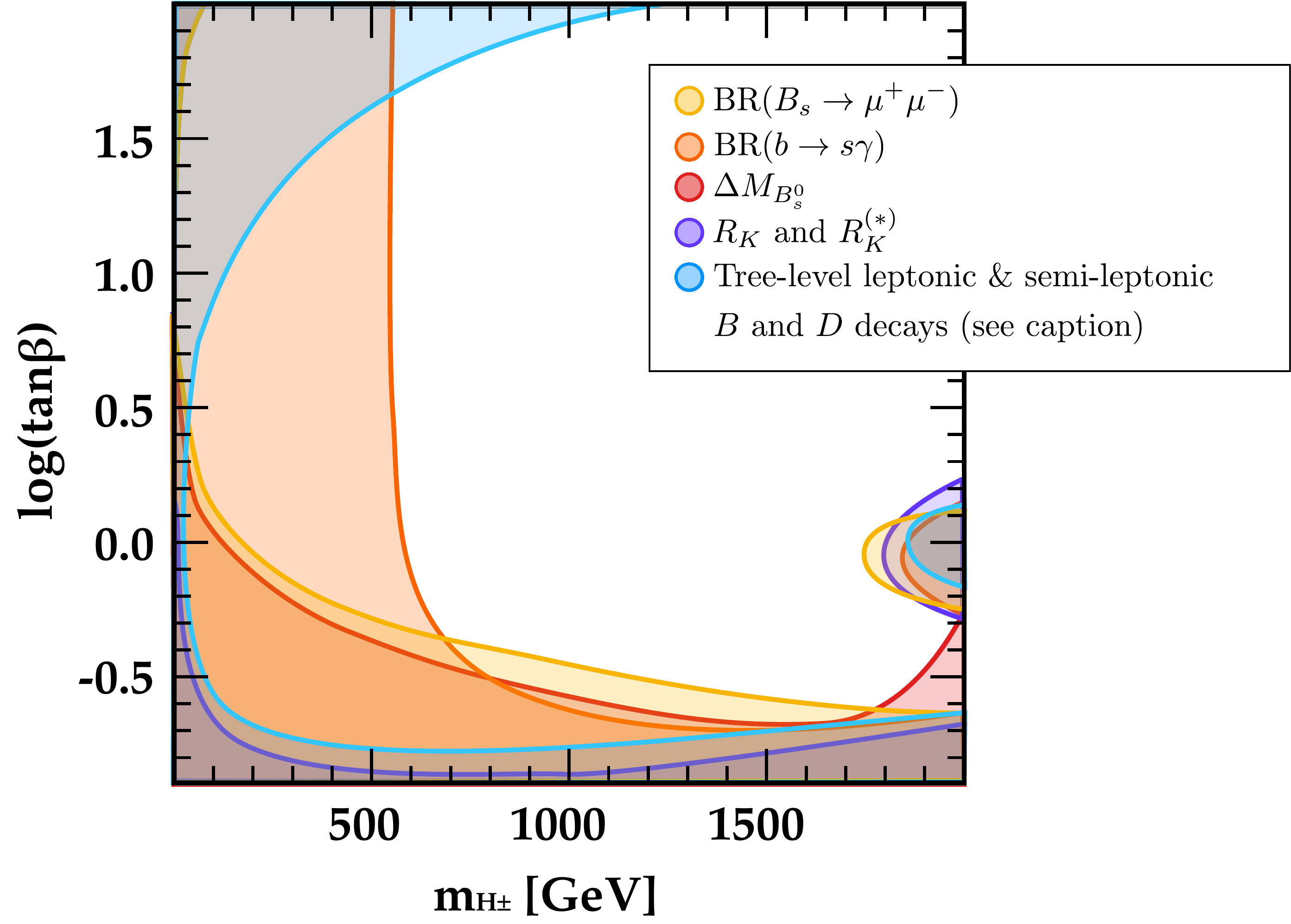}
\caption{Flavour constraints in the plane of $m_{H\pm} - {\rm log}(\tan\beta)$ for type-II 2HDM. We plot the excluded regions at $2\sigma$ with colored shadows detailed in the legend. More specifically, the blue shadow represents tree-level leptonic and semi-leptonic $B$ and $D$ decays: $R_D$, $R_D^{(*)}$, $B \rightarrow \tau \nu, B \rightarrow D \mu \nu, B \rightarrow  D^* \mu \nu$ and $D_s \rightarrow \tau \nu, D_s \rightarrow \mu \nu, D \rightarrow \mu \nu$.}
\label{fig:flavour}       
\end{figure}

In Figure~\ref{fig:flavour}, we show the 2$\sigma$ exclusion regions in the $m_{H^\pm}$ vs tan$\beta$ plane for each flavour likelihood. Fits for each flavour constraint are carried out independently, so there is no pull between the constraints. We see that all of the constraints are in tension with one another.
If we include the effect of measurements of the $B\rightarrow K^*\mu^+\mu^-$ radial components, we find that no region of the 2HDM type-II model parameter space is open at $2\sigma$\footnote{This statement is using the fact that each flavour observable have been fitted independently.}. The $B\rightarrow K^*\mu^+\mu^-$ radial component measurements are anomalous over various energy bins. 
Thus we conclude that the current measurements of $B\rightarrow K^*\mu^+\mu^-$ angular observables cannot be explained in the type-II 2HDM, and we thus ignore their exclusion region in \autoref{fig:flavour}\footnote{The $B\rightarrow K^*\mu^+\mu^-$ angular observables exclude scalar masses above $~500$ GeV, providing a tension with $\text{BR}(b\rightarrow s\gamma)$. When including these in the final analysis we find our heavy scalar masses as significantly pushed downwards.}. For others, $\text{BR}(b\rightarrow s\gamma)$ (orange) disfavours masses below $\sim 500$ GeV as well as low $t_\beta$ values. Low $t_\beta$ is most strongly disfavoured by $\text{BR}(B_s\rightarrow \mu^+\mu^-)$ (yellow) with the lowest value at $2\sigma$ of $t_\beta = 0.4$ near masses of $1.8$ TeV. Past this point the $\Delta_{M_{B^0_s}}$ constraint drives up the lower limit on $t_\beta$. We see that, near our best-fit point for the type-II model, the lower limit imposed on $t_\beta$ sits at around $0.5$ at $2\sigma$.
At large values of $t_\beta$ tree-level leptonic and semi-leptonic $B$ and $D$ decays (blue) come into consideration and push up the lower mass limit on $m_{H\pm}$.

\subsection{Higgs measurements}
\begin{figure}[h]
\centering
\includegraphics[width=6cm,clip]{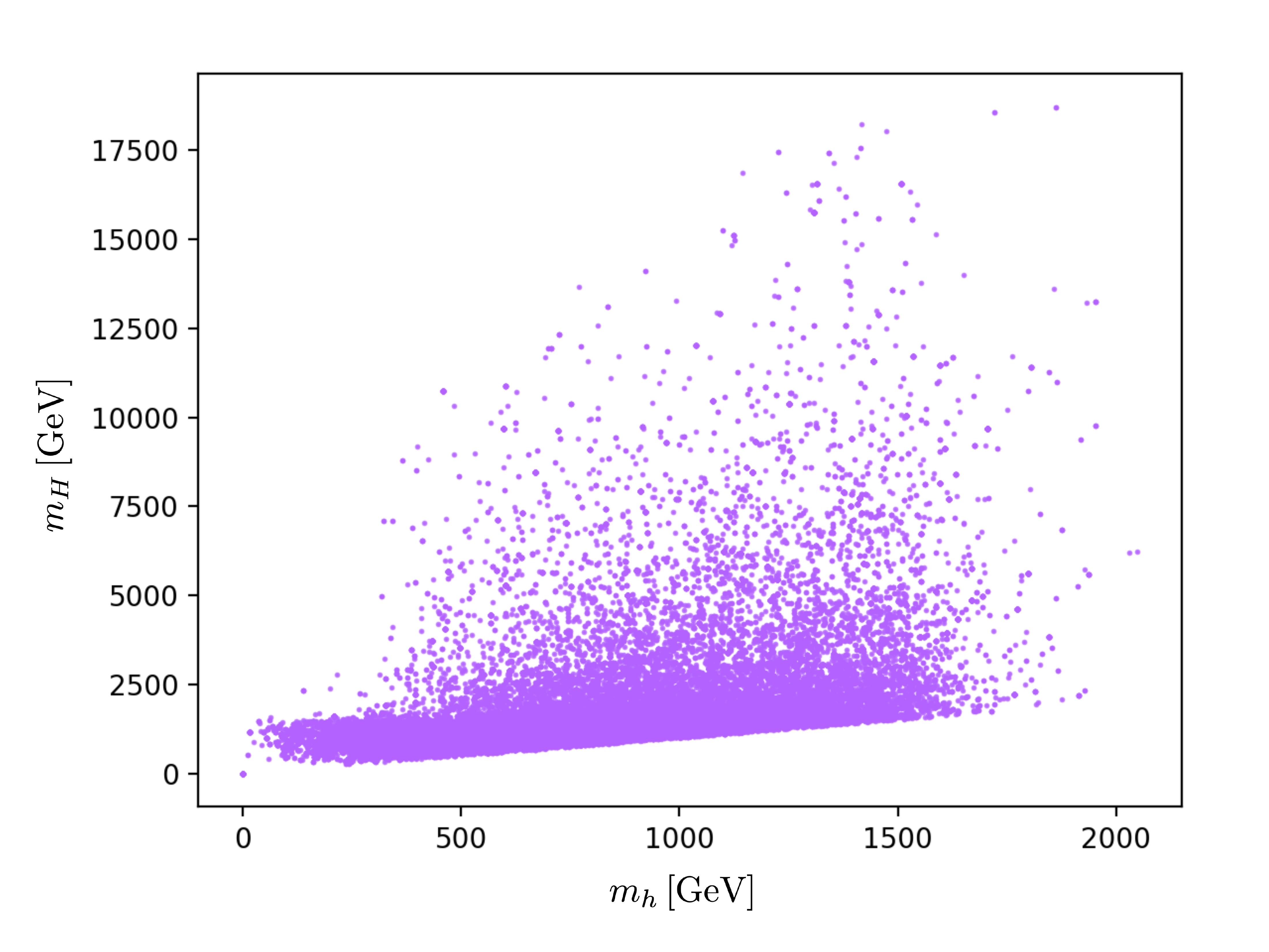}
\includegraphics[width=6cm,clip]{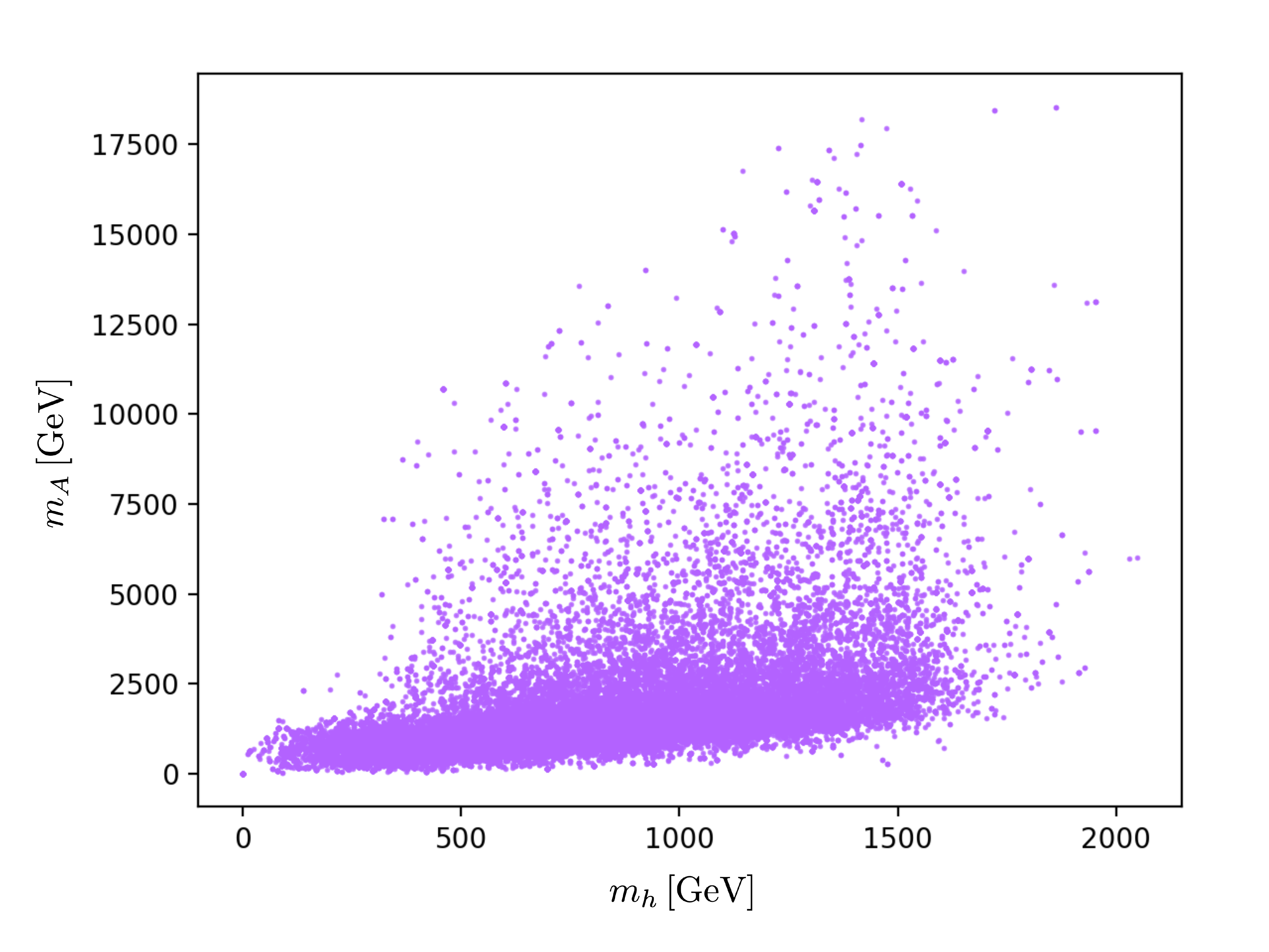}
\caption{Scatter plot in the plane $m_{h, \text{pole}}$ vs $m_{H, \text{pole}}$ \emph{(left)} and $m_{h, \text{pole}}$ vs $m_{A, \text{pole}}$ \emph{(right)} for calculated values when including two-loop corrections. Here we have also included a check on perturbativity to the scalar loop corrections.
}
\label{fig:higgs}       
\end{figure}

For the Higgs sector, all scalars are strongly correlated in mass as expected from the electroweak precision observables fit.
We notice a distinct upper limit on all heavy scalars at around $2$ TeV. This limit appears during spectrum generation with only the SM Higgs boson mass constraint imposed \cite{Su:2019ibd,Su:2019dsf}, as shown in \autoref{fig:higgs} with $m_{h}=125$ GeV. Loop corrections to the light CP-even scalar are of the form $\lambda_i m_\phi$ where $i=3,4,5$ and $m_\phi$ is a term proportional to the mass of the heavy scalars. This relationship implies that, as the heavy scalar masses grow in size, so do the loop corrections to the light scalar (in the case that $\lambda_i$ with $i=3,4,5$ are not close to zero). Loop correction growth eventually saturates to the point where we are no longer able to fit the SM Higgs scalar mass to the pole mass of the light CP-even scalar. In~\autoref{fig:higgs} we present a scatter plot with the pole mass of the light CP-even Higgs scalar against that of the heavy CP-even Higgs scalar on the left and the heavy CP-odd Higgs scalar on the right. In the scatter plot, we also ensure that loop corrections to both scalars remain perturbative. We may read off an upper bound on $m_H$ of $\sim 2$ TeV at $m_h=125$ GeV. The same scatter plot is obtained for each of the heavy scalars.
It is important to mention that this upper limit is not necessarily a limit of the theory but may arise from the form of the loop-order calculation. Specifically, there may exist cancellations to $m_h$ at higher orders that push down the loop-corrections and recover the high mass parameter space.
The fit to an SM-like CP-even scalar is done using \texttt{HiggsSignals} \cite{Bechtle_2014} in the combined likelihood.

A lower limit on $t_\beta$ at $\sim 1$ appears, which we may also attribute to the fit of a SM-like CP-even scalar. Low values of $t_\beta$ are hard to reach when fixing the mass $m_{h}=125$ GeV. At tree-level, with trivial $\lambda_3,\lambda_4$ and $\lambda_5$ we find $m_h^2 = \frac{m_{12}^2}{t_\beta} + v^2 s^2_\beta\lambda_2$ and so a small $t_\beta$ must be cancelled by a small $m_{12}^2$ restricting the allowed parameter space (stability of the potential requires $\lambda_2>0$).

\begin{figure}[h]
\centering
\includegraphics[width=6cm]{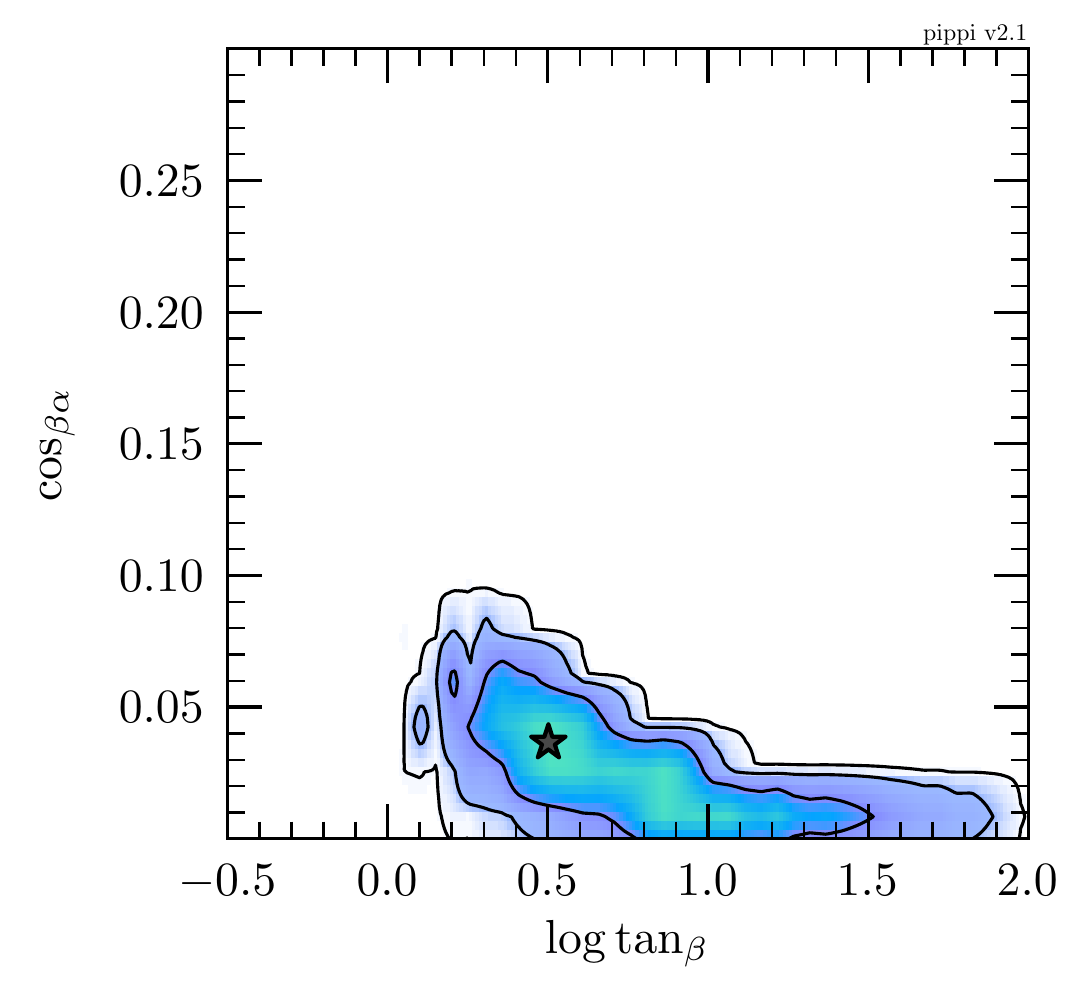}
\includegraphics[width=6cm]{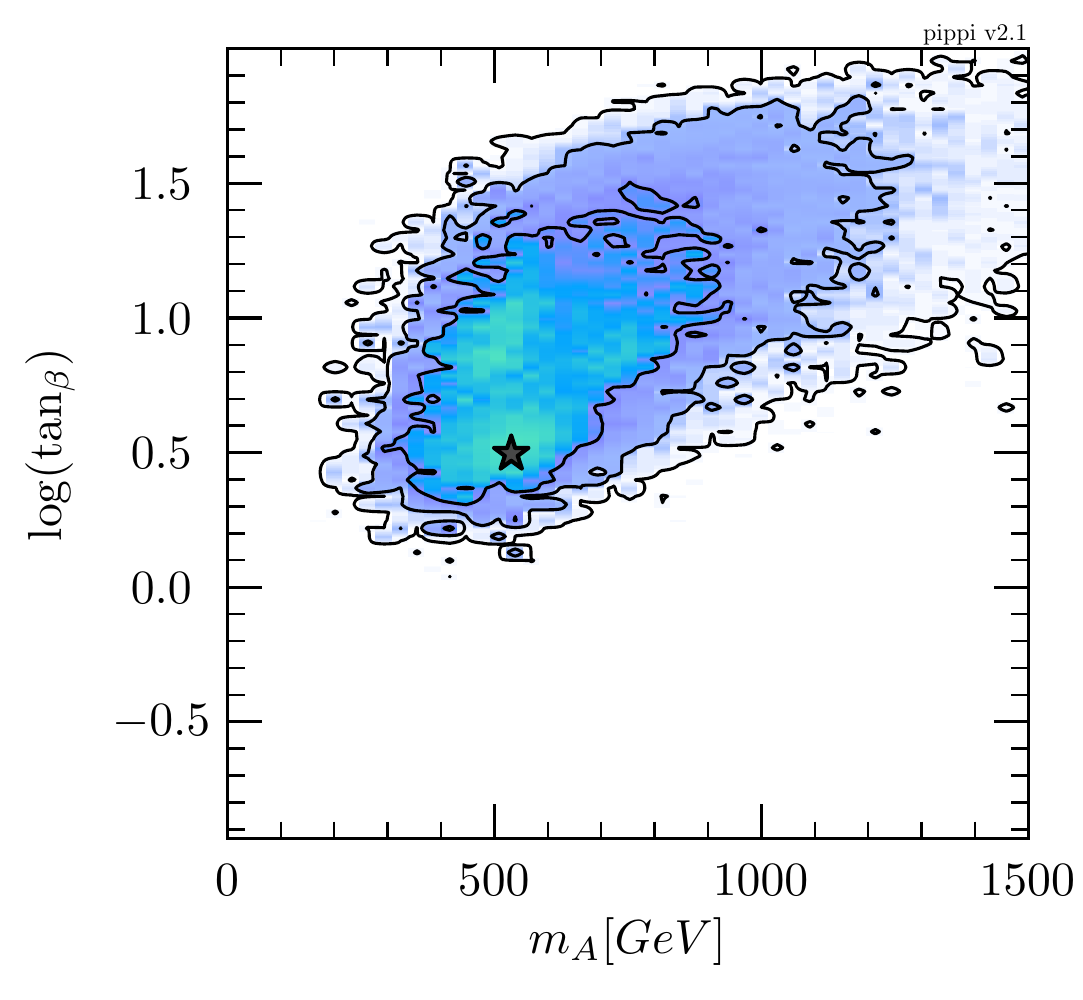}
\caption{ 2D profile likelihood distributions with all constraints switched on in the planes $c_{\beta\alpha}$ vs $t_\beta$ \emph{(left)} and $m_{A, \text{pole}}$ vs $t_\beta$ \emph{(right)}. The solid lines represent $1,2$ and $3\, \sigma$ regions and the best-fit points are plotted as black stars.
}
\label{fig:total}       
\end{figure}

\subsection{Total Global Fit Results from \texttt{GAMBIT}}
For the final global fits including all constraints discussed, 
the interesting results are the distributions of the mixing angles $\alpha$ and $\beta$ as well as that of the heavy Higgs mass. We plot the 2D profile likelihood distributions in the plane $c_{\beta\alpha}$ vs $t_\beta$ and $m_{A, \text{pole}}$ vs $t_\beta$ in \autoref{fig:total} on the left and right panels respectively. In the left panel, it is revealing to plot $c_{\beta\alpha}$ as we are close to the alignment limit that $c_{\beta\alpha}=0$ and it is difficult to read off $s_{\beta\alpha}$ values here.
The best-fit points in each plot are shown with a black star.
In the right panel, we plot pole masses since running masses are never much larger or smaller due to the perturbativity on the loop-correction constraint. Our study shows the agreement between the running mass and pole mass regions of $m_{H,A,h}$ in the type-II model. As a general summary, we can get the typical allowed range $t_\beta \in (1, 50), m_A \in (300, 1000) ~\rm{GeV}$.

\section{Conclusion}
\label{sec:con}
In this work,  we presented \emph{preliminary} results of a global fit of the type-II 2HDM with the tool \texttt{GAMBIT}. We investigated the effect of theoretical constraints (unitarity, perturbativity and vacuum stability), Higgs searches at colliders, electroweak physics and flavour constraints individually, as well as displaying the final results including all constraints.  We found that the typically allowed region is $t_\beta \in (1, 50), \, m_A \in (300, 1000) ~\rm{GeV}$, where the mass upper limit comes from the loop corrected SM-like Higgs mass constraints.
\bibliography{refs}

\end{document}